\def\year{2016}\relax
\documentclass[letterpaper]{article}

\usepackage{aaai16}
\usepackage{times}
\usepackage{helvet}
\usepackage{courier}
\usepackage{graphicx}
\usepackage{amsfonts}
\usepackage{amsmath}
\usepackage{booktabs}
\usepackage{color}
\usepackage{subcaption}
\usepackage{hyperref}

\begin{document}
 
\title{Distinguishing between Topical and Non-topical Information Diffusion Mechanisms in Social Media}

\author{
Przemyslaw A. Grabowicz\\
MPI-SWS, Germany
\And
Niloy Ganguly\\
IIT Kharagpur, India
\And
Krishna P. Gummadi\\
MPI-SWS, Germany
}

\maketitle

\begin{abstract}

A number of recent studies of information diffusion in social media, both empirical and theoretical, have been inspired by viral propagation models derived from epidemiology. These studies model the propagation of memes, i.e., pieces of information, between users in a social network similarly to the way diseases spread in human society. Importantly, one would expect a meme to spread in a social network amongst the people who are interested in the topic of that meme. Yet, the importance of topicality for information diffusion has been less explored in the literature. 

Here, we study empirical data about two different types of \textit{memes} (hashtags and URLs) spreading through the Twitter's online social network. For every meme, we infer its topics and for every user, we infer her topical interests. To analyze the impact of such topics on the propagation of memes, we introduce a novel theoretical framework of information diffusion. Our analysis identifies two distinct mechanisms, namely topical and non-topical, of information diffusion. The non-topical information diffusion resembles disease spreading as in \textit{simple contagion}. In contrast, the topical information diffusion happens between users who are topically aligned with the information and has characteristics of \textit{complex contagion}. Non-topical memes spread broadly among all users and end up being relatively popular. Topical memes spread narrowly among users who have interests topically aligned with them and are diffused more readily after multiple exposures. Our results show that the topicality of memes and users' interests are essential for understanding and predicting information diffusion.

\end{abstract}

\section{Introduction}
\label{sec:intro}
The process of information diffusion is crucial for viral marketing \cite{Leskovec2007Dynamics}, spreading of political movements \cite{Strang1998Diffusion} and innovations \cite{Rogers1962Diffusion}, and opinion formation \cite{Holley1975Ergodic}. 
Numerous studies have shown that information and behaviors propagate virally over online social networks via \textit{social exposures}, defined as the number of friends or acquaintances of a user that have exposed the meme to her~\cite{Backstrom2006Group,Cosley2010Sequential,Crandall2008Feedback,Gomez-Rodriguez2014Quantifying,Romero2011Differences}. As such, the phenomenon of information diffusion in social networks is predominantly modeled by building parallels with the viral spread of diseases \cite{Kempe2003Maximizing}, or by using graph-based methods \cite{Granovetter1978Threshold,Kempe2003Maximizing}.
However, information diffusion involves intelligent individuals exhibiting complex behaviors, hence one can expect that it includes mechanisms beyond viral spreading via social exposures.
An important mechanism that may drive such spread is the topical alignment 
between the interests of individuals and the pieces of information, which in the language of epidemiology is expressed as the genetic susceptibility to a certain disease~\cite{Burgner2006Genetic}.

However, it is difficult to empirically observe this phenomenon for information diffusion.
First, large-scale datasets of information that are not influenced by recommender systems and contain enough information for topic modeling are scarce. 
Second, topic models have matured and have become viable for large datasets only in recent years.
We overcome these issues thanks to the increasing availability of data characterizing users and content in online social networks \cite{Cha2010Measuring} and the development of reliable topic modeling methods \cite{Blei2003Latent,McCallum2002MALLET:}.
Third, information diffusion within a given system may be caused by the exposures external to that system, which we are not aware of. Thus, it is challenging to isolate the effect of the external exposures from the effect of the internal social exposures on the information diffusion~\cite{Bakshy2012role}.

Here, to separate internal exposures from external exposures and to analyze the effect of topics on information diffusion, we introduce a theoretical framework that models topics of memes and users as latent variables and the probability of diffusion as a function of these variables.
Diffusion processes are usually described statistically with the so-called \textit{adoption probability}~\cite{Backstrom2006Group,Cosley2010Sequential,Crandall2008Feedback,Gomez-Rodriguez2014Quantifying,Romero2011Differences}, defined as the probability of adopting a meme by a person who has been exposed to this meme by exactly $\kappa$ social exposures.
Our theoretical framework allows us to extend the definition of adoption probability by accounting for the topical alignment between a meme and interests of each user and to estimate the information diffusion from both internal and external sources.

Furthermore, we hypothesize that there exist two different mechanisms of information propagation, namely, a topical and non-topical information diffusion.
Under this hypothesis, topical information would start to spread from the enthusiasts or experts of that topical domain. Then, it would diffuse narrowly to other members of this domain. Non-topical information would not be limited by its topical domain. If such hypothesis is correct, then one can expect that topical and non-topical information diffusion have different properties. 
Here, we aim to identify characteristics specific to each of these two mechanisms. 
Finally, having the two mechanisms identified and characterized, it is crucial to relate them to existing theories of information diffusion. The sociological theory of complex contagion suggests that multiple social exposures are important for the diffusion of novel and controversial information~\cite{Centola2007Complex}, while the theory of simple contagion posits viral spreading after the initial exposure. How these theories relate to topical and non-topical information diffusion?
These are some of the questions that we address in this study.

\subsection{Present work}

Note that, to understand the importance of topicality for information diffusion, we require (i) a closed social system that does not implement any personalized recommendation algorithm\footnote{At the time of data collection, all follower links were created by users themselves without the influence of any personalized recommendation algorithm.}, (ii) a comprehensive and independent annotation of topicality of users and diffused pieces of information, and (iii) a theoretical framework that is able to estimate the effect of internal and external social exposures on information diffusion and to quantify topicality of memes and users. 


To this end, we analyze a dataset from Twitter containing all public tweets that were messaged during the observation period of three months in mid-2009 and a full snapshot of the follower graph from the end of that period. 
During that time, there were no personalized recommendations in Twitter and no other mechanisms than follower links to subscribe to other users. Hence, we are able to measure information diffusion in a {\em natural setting}, namely in an online social network that is hardly affected by any recommender system. 
We consider two distinct types of information that diffuse over the follower graph, namely hashtags and URLs, which for simplicity we jointly call \textit{memes}. We say that a user \textit{diffuses} or \textit{adopts} a meme if she writes it in one of her tweets.

Next, we split the dataset chronologically into two parts (see Figure~\ref{fig:timeline}). We process the prior part to infer the topics of memes and users' interests, whereas we use the latter part to perform an independent analysis of information diffusion. More specifically, using nouns extracted from past tweets we infer with an unsupervised method, namely with Latent Dirichlet Allocation, i.e., LDA~\cite{Blei2003Latent}, the topical distributions of active users and popular memes (Section~\ref{sec:topics}). 
As such, in our theoretical framework, topics are mixtures of words that frequently appear together. Each user and meme is represented by a topical distribution over a constant number of topics. To evaluate how topically narrow are memes and users' interests, we compute the entropy of their corresponding topical distributions.
Thanks to this procedure, we distinguish between \textit{topical} and \textit{non-topical} memes and users. We say that a meme is topical if it is strongly associated with a topic, whereas a non-topical meme is weakly and equally associated with all topics. 
Importantly, we define the topical alignment between a meme and interests of a user as a cosine similarity of their respective topical distributions. 
We exploit this topical alignment to compute the probability of adoptions triggered either via sources \textit{internal} or \textit{external} to Twitter (Section~\ref{sec:framework}). We find that topical memes are likely to be topically aligned with the interests of people who adopt them, either via internal or external exposures.
We leverage out theoretical framework further, finding that:
\begin{itemize}

\item A considerable fraction of memes are topical. (Section~\ref{sec:topics})

\item We show that adoptions of topical memes are easier to predict than the adoptions of non-topical memes. (Section~\ref{sec:adoption-curves-memes})

\item For the topical information diffusion, the probability of adoption steadily increases with the topical alignment between user's interests and a meme, whereas for non-topical information diffusion the probability stays flat or increases considerably less. (Section~\ref{sec:adoption-curves-memes})

\item Topical information diffusion profits more than non-topical diffusion from multiple exposures to a meme; thus, it resembles complex contagion. We capture it by introducing a novel metric of spreading persistence. (Section~\ref{sec:adoption-curves-users})

\item A large fraction of adoptions happens via seed adopters exposed to external sources. The seed adopters of topical memes are relatively more aligned with the meme than other adopters and the seed adopters of non-topical memes. (Section~\ref{sec:seeds})

\item The diffusion of URLs is about twice more intense via external rather than internal sources, whereas the diffusion of hashtags has similar intensity both for external and internal exposures. (Section~\ref{sec:seeds})

\end{itemize}

\subsection{Related work}
\label{sec:rw}

Some of our results can be interpreted with the sociological theory of \textit{complex contagion}~\cite{Centola2010Spread,Centola2007Complex}. This theory posits that ``high-risk'', ``novel'', ``unproven'', or ``controversial'' information spreads via ``wide bridges'', namely via multiple social exposures or strong ties. As we show later, topical memes convey such information.
Romero et al. report that successive social exposures to political hashtags tend to increase adoption probability more than multiple social exposures to idioms~\cite{Romero2011Differences}. The authors explain these diffusion classes with complex contagion mechanism.
Our work generalizes their analysis and methodology, in the sense that we do not focus on a set of specific topics (e.g., politics and idioms), but instead we study a quantifiable and robust topical space, which represents such specific topics, yielding consistent results. 
Our findings suggest that topical information diffusion is related to complex contagion. 

It is intuitive that human behavior is largely related to multi-dimensional personal characteristics of people, such as topical interests or private traits \cite{Kosinski2013Private}. 
For instance, some of the recently proposed models suggest that latent factors impact information diffusion \cite{Barbieri2012Topic,Du2013Uncover} and measure topical social influence \cite{Weng2010TwitterRank:,Bi2014Scalable}. 
Moreover, recommender systems exploiting the premise of latent factors predicting content consumption are widely deployed \cite{Ricci2011Introduction}.
However, empirical studies of information diffusion in social media tend to focus either on predictions of the size of social cascades \cite{Ma2013Predicting,Weng2014Topicality}, analysis of the structure of the cascades \cite{Goel2012structure}, or detecting influential spreaders \cite{Cha2010Measuring}, which are macroscopic and mesoscopic properties of information diffusion.
A few empirical works study the information diffusion at the microscopic level. These studies explore, for instance, the classes of information diffusion for specific topics \cite{Romero2011Differences}, the impact of activity similarity on the probability of interaction~\cite{Crandall2008Feedback}, and the predictiveness of limited content-based features for future adoptions~\cite{Yang2012We,Yuan2016Who}. 
Thus, to our knowledge, our work is the first large-scale empirical study systematically analyzing the relation between topicality and mechanisms of information diffusion and the extent to which topical alignments affect information diffusion.

\section{Dataset}
\label{sec:dataset}


\begin{figure}[b]
\centering
\includegraphics[width=8cm]{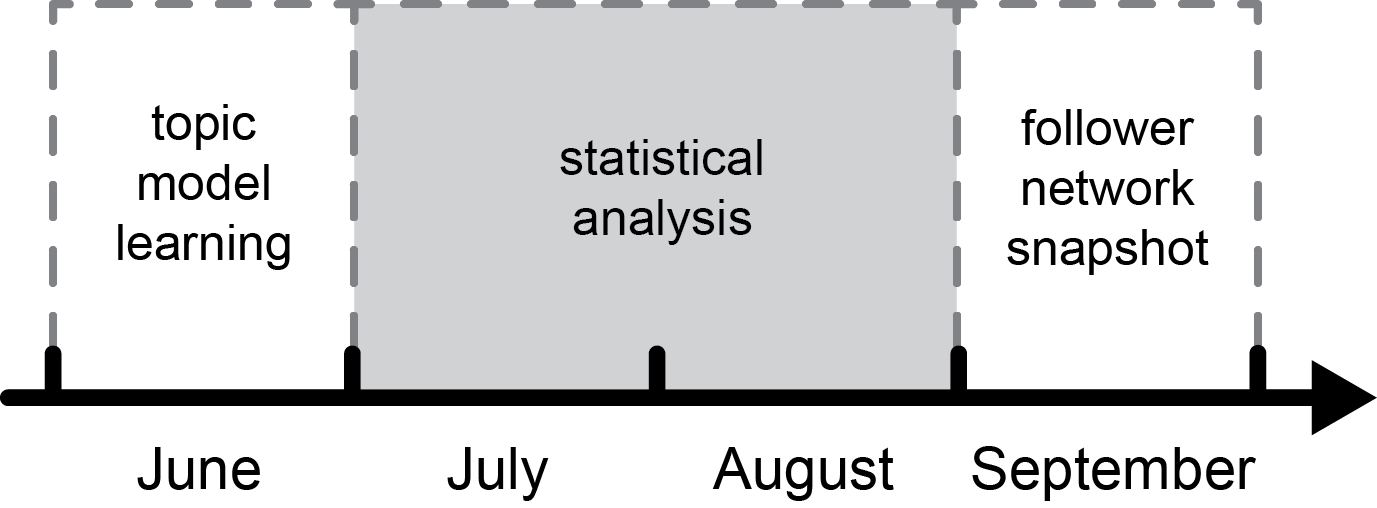}
\caption{Timeline of the dataset illustrating the design of the study. The gray area marks the period of analysis.}
\label{fig:timeline}
\end{figure}

We analyze a dataset from Twitter that is nearly complete and contains all public tweets produced by users until September 2009 and a snapshot of the social graph crawled in September 2009 \cite{Cha2010Measuring}. 
In Twitter, a user can post short messages, called tweets, that are delivered to all her \textit{followers}, namely all the users who follow her in the follower graph. (The users who are followed by the user are her \textit{followees}.) Such messages can contain memes, so the memes can diffuse over the follower graph. 
Importantly, at the time period when our dataset was gathered, there were no other ways than the follower links to directly receive content from other users.
We study the diffusion of two types of memes: hashtags and URLs. We say that a user becomes an \textit{adopter} of a meme and starts to spread it once she posts it at least in one tweet. Naturally, at the moment of posting the meme, the adopter becomes an \textit{exposer} of the meme to all her followers. 

We analyze adoptions that happened during the period of three months, from June to August 2009. During this time users posted nearly a billion of tweets. We consider \textit{only} the memes that emerged during that period, specifically during June. In other words, we do not consider memes that were born before that period. This constitutes a difference from the prior works \cite{Gomez-Rodriguez2014Quantifying}, which tracked also the memes that emerged before the observation period, hence underestimating the number of social exposures.\footnote{The user could have been exposed to the meme before the observation period. If this happens and one does not have knowledge about it, then the real social exposure is higher than the one estimated based solely on the observation period. Our dataset covers all previous months apart from the three-months long period of observation. Thus, we can accurately estimate the real number of exposures.} We focus on memes that were used predominantly in English tweets. Specifically, we analyze the memes for which at least $90\%$ of tweets were in English. To detect the language of tweets we use a state-of-the-art language detection tool for Twitter data \cite{Nakatani2012Language}. Furthermore, we report results only for the memes that have been adopted by at least $100$ unique users.\footnote{We obtain qualitatively the same results for lower thresholds.} Overall, we analyze $704$ hashtags adopted $480$k times by a total of $330$k unique users and $3745$ URLs adopted $1.2$m times by a total of $610$k unique users. 

We divide the three-months long dataset into two parts. The first part consists of all tweets from June, while the second part consists of all tweets from July and August (Figure~\ref{fig:timeline}). We use the first part of the dataset solely to infer topical distributions of users' interests and memes emerging during that month. This partition allows us to keep the topic modeling independent from the statistical analysis, which is performed solely on the second part of the dataset.

\begin{table}[t]
\resizebox{0.47\textwidth}{!}{
\centering
\begin{tabular}{lp{6.66cm}}
\toprule
Iran election & mousavi, basij, khamenei, ppl, protesters, rally\\
British election & bbc, london, bnp, news, brown, bit\\
Well-being & success, happiness, einstein, albert, john, failure\\
Viral marketing & marketing, business, twitter, media, blog, site\\
Smart devices & iphone, apple, app, mac, macbook, wwdc\\
\bottomrule
\end{tabular}
}
\caption{An example of five topics detected by LDA and words associated with each of them. The words are ordered by their relevance to the topic. (Words ``albert'' and ``einstein'' are associated with one of the topics because his sayings are quoted several times within that topic.)}
\label{tab:example_topics}
\end{table}

\begin{table*}[t]
\resizebox{\textwidth}{!}{
\centering
\begin{tabular}{c|c|p{20cm}}

\multicolumn{3}{c}{\Large{The most topical hashtags}}\\
\toprule
Hashtag	&	Topic	&	Examples of tweets hand-picked from 30 chronological tweets from the beginning of June 2009\\
\midrule

\#ashraf	&	Iran election	& 

1. canadian friends of a democratic iran urge us president to ensure ashraf residents treated according international law \#iran \#ashraf \#rajavi
\newline
2. suppressive forces heavily present in streets of iran capital to prevent protests \#iran \#iranelection \#tehran \#neda \#rajavi \#ashraf
\newline
3. how u can help iranians http://tinyurl.com/m8fun4 \#iranelection \#tehran \#azadi \#ashraf \#rajavi \#neda \#basij \#iran \#gr88 rt rt rt
\newline
\\

\#electionstudio	&	British election	&

1. bnp is a left wing party. so were the german national socialists. \#electionstudio
\newline
2. \#electionstudio this bnp guy is a lunatic frankly.  indigenous population?  if it weren't for immigration britain would not exist at all.
\newline
3. \#electionstudio i got very excited last night when purnell resigned but now can't see a leadership challenge coming from anywhere
\newline
\\

\#vene2ia	&	Well-being	&	

1. without freedom of choice there is no creativity~kirk, star trek \#vene2ia
\newline
2. it wasn't until quite late in life that i discovered how easy it is to say "i don't know!" ~somerset maugham \#vene2ia (via @vene2ia)
\newline
3. rt @vene2ia: children are the living messages we send to a time we will not see~john w. whitehead \#vene2ia
\newline
\\

\#smum	&	Viral marketing	&	
1. learn how to blog your way to success at social media university milwaukee http://ow.ly/g83u (last day for early bird pricing) \#smum
\newline
2. lrt @philgerb: rt @dnewman:great article on benefits of audience tweeting during your presentation http://bit.ly/rdgsl good 2 know 4 \#smum!
\newline
3. l@backseatgaffer glad you found it. you can sign up even if you're not going to \#smum. details: http://bit.ly/3qyhot reg: http://bit.ly/zvpeh
\newline
\\

\#mytouch	&	Smart devices	&

1. mytouch3g ordering starts tomorrow, very excited \#tmobile \#android \#mytouch3g
\newline
2. im thinking of getting the tmobile \#mytouch phone, i want to know, what do you all think of the phone? should i get it?
\newline	
3. if you're using a \#bold, \#curve, \#g1, \#mytouch or an \#iphone you should download @loopt an amazing lbs networking app
\\
\bottomrule

\multicolumn{3}{c}{}\\
\multicolumn{3}{c}{\Large{The least topical hashtags}}\\
\toprule
Hashtag	&	Topic	&	Examples of tweets hand-picked from 30 chronological tweets from the beginning of June 2009\\
\midrule

\#uknowyouugly	&	-	&	

1. \#uknowyouugly when you're at the zoo and the monkeys run from you
\newline
2. \#uknowyouugly when u gota get fresh to go to the grocery store
\newline
3. \#uknowyouugly when they ban you from the mac store
\newline
\\

\#unala2009	&	-	&	

1. getting ready for my first unconference... the first one might as well be a big one! very excited. \#ala2009 \#unala2009
\newline
2. got new business cards designed by my son. unconference tag \#unala2009 per conference wiki. ala facebook meetup: http://tinyurl.com/loayhl
\newline
3. just for clarification, friday's unconference is at the chicago hilton, not the palmer house hilton. hth \#ala2009 \#unala2009
\newline
\\

\#skiphop	&	-	&	

1. i just entered to \#win a \#skiphop bag with a \#beautyfix kit in the \#99daysofsummer \#contest at \#modernmom! http://bit.ly/wetvi
\newline
2. modernmom i just entered to \#win a \#skiphop bag with a \#beautyfix kit in the \#99daysofsummer \#contest at \#modernmom! http://bit.ly/p29uz
\newline
3. i just entered to \#win a \#skiphop bag with a \#beautyfix kit in the \#99daysofsummer \#contest at \#modernmom! http://bit.ly/7ap8x \#trackle
\newline
\\

\#sexisbetter	&	-	&

1. \#sexisbetter wen your drunk
\newline
2. \#sexisbetter with a scorpio! (guess what sign i am) ;-)
\newline
3. \#sexisbetter when its anywhere but a bed
\newline
\\

\#besthead	&	-	&	

1. \#besthead when dey don't expect it baq.. lol
\newline
2. \#besthead is when u black out and just pass out now dats da best
\newline
3. \#besthead was in the morning right after she brushed her teeth,she went to work!! u heard me (master p voice)
\\

\bottomrule

\end{tabular}
}
\caption{The top five most and least topical hashtags. For each hashtag, we show three exemplary tweets and a topic corresponding to this hashtag. The least topical hashtags are not associated with a specific topic. In this list, we omit all hashtags that co-appear with the listed hashtags.}
\label{tab:example_memes}
\end{table*}

\section{Topicality of memes and users}
\label{sec:topics}

Here, we describe how we obtain the topical distributions of memes and users' interests. Then, we introduce the notion of topicality to distinguish between topical and non-topical memes and user's interests based on their respective topical distributions. Finally, we introduce the topical alignment between a meme and user's interests.

\subsection{Topical distributions}

We define topical distributions for each meme and user based on all tweets posted in June. First, we filter out the tweets that are not written in English. Second, we extract from remaining tweets all nouns and proper nouns using a state-of-the-art POS-tagger for Twitter data \cite{Owoputi2013Improved}. We choose nouns and proper nouns, because they encapsulate the semantics of tweets.
We represent each meme as a bag of unique nouns that appeared in tweets containing this meme and each user as a bag of unique nouns that were tweeted by this user. Next, we model the topics of users and memes with LDA performed for $100$ topics and $1000$ iterations \cite{McCallum2002MALLET:}.\footnote{We also tested lower and higher numbers of topics, finding similar results both qualitatively and quantitatively.} Hence, the topics of user's interests and a meme are represented by distributions over $100$ topics.
Examples of topics inferred with this method are shown in Table~{\ref{tab:example_topics}. Note that each such topic gathers related terms. We discuss these topics further in the following paragraphs.

\begin{figure}[btp]
\begin{picture}(100,100)
\centering
\put(0,0){
\includegraphics[width=0.23\textwidth]{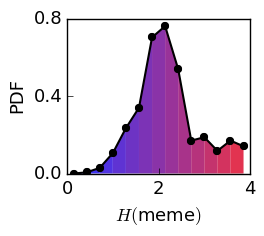}
\includegraphics[width=0.23\textwidth]{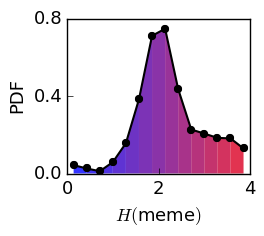}
}
\put(55,97){hashtags}\put(180,97){URLs}
\put(35,85){A}\put(155,85){B}
\put(40,50){\rotatebox{90}{\textcolor{blue}{topical}}}
\put(100,45){\rotatebox{90}{\textcolor{red}{non-topical}}}
\end{picture}\\
\begin{picture}(100,100)
\centering
\put(0,0){
\includegraphics[width=0.23\textwidth]{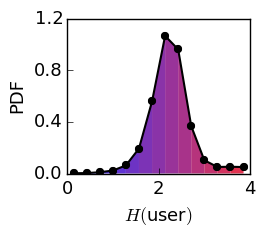}
\includegraphics[width=0.23\textwidth]{pub-distros-pdf-expentropy-peruser-static_tot_user_redpart-notrans_adoption-properties16_hashtag-uall-allflt_joints-mallda2-tmsplit-avg-dim100_emerge-muut100_std}
}
\put(35,85){C}\put(155,85){D}
\end{picture}
\caption{The distribution of entropy of topical distributions of (A-B) memes and (C-D) user's interests. The memes or Users' interests with the lowest entropy are the most topical, while the memes with the highest entropy are the least topical.}
\label{fig:distro_entropy}
\end{figure}

\subsection{Topicality}

To measure how much topical a meme or a user's interests are, we compute the entropy $H$ of their respective topical distributions. The higher the entropy, the less topical the meme or the user are, and vice versa. 
It follows that the most topical would be a meme that is associated only with a single topic, whereas the least topical would be a meme that is equally associated with all topics, i.e., has a uniform distribution over all topics. Most of memes is between these two extreme examples.
The distributions of entropy resemble a normal distribution (Figure~\ref{fig:distro_entropy}). We refer to the $25\%$ most topical memes and users as \textit{topical}, whereas to the $25\%$ least topical as \textit{non-topical}. We establish this distinction for the sake of clarity, however, other thresholds (e.g., at $50\%$) give qualitatively the same results. In the remainder, we present results both for topical and non-topical memes and users.

We list the most and the least topical memes in Table~\ref{tab:example_memes}. For instance, hashtags \#ashraf and \#electionstudio correspond to political elections in Iran and Great Britain, \#vene2ia represents a motivational website, \#smum is about a newly-established viral marketing school, and \#mytouch corresponds to a newly released smartphone. 
Note these hashtags correspond closely to the categories of information that is predicted to spread via complex contagion, namely novel, unproven, high-risk, and controversial information.

On the contrary, the least topical hashtags are mostly Twitter idioms and neologism that can be understood directly from their names, i.e., \#uknowyouugly, \#sexisbetter, \#besthead, or are related to a popular lottery, i.e., \#skiphop. These idioms have a generic entertainment value. The hashtag \#unala2009 corresponds to an un-conference, which is a conference without agenda. 

Strikingly, the distinction between topical, e.g., political, hashtags versus idioms corresponds very closely to the categories of hashtags exhibiting complex and simple contagions described in~\cite{Romero2011Differences} (see Section~\ref{sec:rw}). This indicates that the topical information diffusion is related to complex contagion.

Finally, we note that topical memes are $(42 \pm 2)\% $ more likely to be adopted by a topical user than by a non-topical user. This is in a sense the tip of the iceberg, because here we do not take into account topical alignments between memes and users, which increase the probability of adoption much further.

\subsection{The topical alignment between a meme and a user}

Once we have the topical distributions of memes and users, we measure the topical alignment between a user and a meme, denoted as $S(\text{user},\text{meme})$, by computing the cosine similarity of the corresponding topical distributions. The topical alignment assumes values from $0$ to $1$, where $1$ corresponds to the maximum possible alignment.
The alignment can be $1$ both for topical and non-topical memes and users. 
Even if a meme is perfectly non-topical (has a uniform distribution over topics), its topical alignment with a user that has interests distributed uniformly over topics will be $1$.

\label{sec:adoption-curves}

\section{The theory of adoption probability}
\label{sec:framework}

In this section, we formally introduce adoption probability and extend its definition to cover topical alignment.
To this end, we define the fundamental random events corresponding to the adoptions and exposures of memes.
Then, we exploit these events to define the adoption probability as a function of social exposure and topical alignment. Also, we show that the two channels of social influence, namely influence internal and external to Twitter, can be separated within our model of information diffusion. We close this section by introducing a new metric of spreading persistence that quantifies the effect of multiple exposures on the probability of adoption. 
The methods and definitions that we introduce here can be used for studying information diffusion in various systems. In the remainder of this study, we apply this theoretical framework to our Twitter dataset.

\subsection{Extending the definition of adoption probability}

One can reason about information diffusion by defining \textit{exposure event} as a tuple (user, meme, exposers, adoption state) that represents a user being exposed by a given set of exposers to a meme that she has not adopted yet. 
Naturally, the set of exposers may be empty, if the user has not been exposed to the given meme yet by anybody. The set of exposers can only grow in time, as more people adopt the given meme.
The adoption state tells us if the meme will be adopted \textit{before} the appearance of a new exposer. If it gets adopted, then we call such exposure event as an \textit{adoption event}, to distinguish it from all other exposure events.
Note that each exposure event can be characterized by the topical alignment between a meme and a user.
Some users in our dataset do not have topical distributions defined. Thus, we filter out all exposure events for such users.

Next we focus on exposure events with $\kappa$ exposers.
Namely, we say that a user is $\kappa$-exposed to a meme at a certain time, if $\kappa$ distinct users that she follows adopted the meme prior to this time and she has not adopted the meme yet. 
The probability of adoption is defined \cite{Cosley2010Sequential}, as the probability that a $\kappa$-exposed user will adopt the meme before getting $(\kappa+1)$-exposed to it, namely
\begin{equation}
P_\text{a}(\kappa) = N_a(\kappa)/N_e(\kappa),
\label{eq:adopt_prob_k}
\end{equation}
where $N_e(\kappa)$ is the number of users who were $\kappa$-exposed to the information, and $N_a(\kappa)$ is the number of users that were $\kappa$-exposed and adopted the meme before becoming $(\kappa + 1)$-exposed.
In other words, $N_e(\kappa)$ and $N_a(\kappa)$ are the numbers of exposure events and adoption events with $\kappa$ exposers, respectively.\footnote{To decrease the impact of noise, we sum up the numbers of events over all memes and then compute $P_\text{a}(\kappa)$. Thus, in comparison with the traditional definition of adoption probability, we weight popular hashtags proportionally more.}

We extend the above standard definition of adoption probability to include a topical alignment. 
Namely, we define the conditional probability that a $\kappa$-exposed user will adopt the meme before getting $(\kappa+1)$-exposed when the topical alignment is $S$, namely 
\begin{equation}
P_\text{a}(\kappa,S) = N_a(\kappa,S)/N_e(\kappa,S),
\label{eq:adopt_prob_2d}
\end{equation}
where $N_e(\kappa,S)$ and $N_a(\kappa,S)$ are the numbers of exposure events and adoption events with the social exposure $\kappa$ and the topical alignment $S$.
One can obtain one-dimensional adoption probabilities $P_\text{a}(\kappa)$ or $P_\text{a}(S)$ by integrating $P_\text{a}(\kappa,S)$ over $\kappa$ or $S$, respectively, which gives
\begin{equation}
P_\text{a}(S) = N_a(S)/N_e(S).
\label{eq:adopt_prob_s}
\end{equation} 
The adoption probability versus social exposure $P_\text{a}(\kappa)$ has been measured in multiple studies. In contrast, to our knowledge, the adoption probability versus a topical alignment $P_\text{a}(S)$ is measured for the first time in this study.

\subsection{Diffusion via internal and external sources}

Information can spread from sources that are internal or external to a given system (in our case, to Twitter). In general, it is difficult to distinguish between internal and external exposures and to estimate their impact on information diffusion~\cite{Bakshy2012role}. 
Even if we know that a user follows a person that tweeted some meme and that this person adopted this meme afterwards, we still do not know whether this adoption was caused by the exposures via Twitter followers or via some other external sources. Furthermore, in the next section we show that external adoptions are frequent for information diffusion on Twitter.

Fortunately, our theory of adoption probability allows us to differentiate between information diffusion via internal and external sources, i.e.,
\begin{equation}
P_\text{a}(\kappa^{\text{i}},\kappa^{\text{e}},S) = P_\text{a}^{\text{i}}(\kappa^{\text{i}},S) + P_\text{a}^{\text{e}}(\kappa^{\text{e}},S).
\label{eq:a}
\end{equation} 
In this study, the internal exposures are via the follower network of Twitter and we can estimate the number of such exposures. By contrast, exposures via external sources are opaque to us. In other words, we do not know $\kappa^{\text{e}}$, so we marginalize it out from the last equation obtaining
\begin{equation}
P_\text{a}(\kappa,S) = P_\text{a}^{\text{i}}(\kappa,S) + P_\text{a}^{\text{e}}(S),
\label{eq:b}
\end{equation} 
where $\kappa \equiv \kappa^{\text{i}}$ and $P_\text{a}^{\text{e}}(S) = \int P_\text{a}^{\text{e}}(\kappa^\text{e},S) P(\kappa^\text{e}|
S) d\kappa^\text{e}$. Next, we note that, if no internal exposures are present, i.e., $\kappa^{\text{i}}=0$ and $P_\text{a}^{\text{i}}(0,S)=0$, then the only influence remaining is external, thus
\begin{equation}
P_\text{a}(0,S) = P_\text{a}^{\text{e}}(S).
\label{eq:c}
\end{equation} 
This simple formula shows how to estimate the probability of adoption via external exposures as a function of topical alignment. 
Combining Equations~\ref{eq:b} and~\ref{eq:c} allows us to estimate the \textit{joint adoption probability via internal exposures}, namely
\begin{equation}
P_\text{a}^{\text{i}}(\kappa,S) = P_\text{a}(\kappa,S) - P_\text{a}(0,S).
\label{eq:d}
\end{equation} 
Finally, we compute the marginalized internal adoption probabilities $P_\text{a}^{\text{i}}(\kappa)$ and $P_\text{a}^{\text{i}}(S)$, as follows:
\begin{align}
P_\text{a}^{\text{i}}(S) = P_\text{a}(S)  - P_\text{a}(0,S), \\
P_\text{a}^{\text{i}}(\kappa) = P_\text{a}(\kappa)  - \int P_\text{a}(0,S) P(S|\kappa) dS.
\label{eq:p_i}
\end{align} 
These two formulas subtract the probability of adoption caused by external sources from the probability of any adoptions. Interestingly, Equation~\ref{eq:p_i} subtracts the expected value of adoption probability given that the conditional probability distribution of similarity is $P(S|\kappa)$. Intuitively, even if an individual is exposed to a meme on Twitter via $\kappa>0$ internal social exposures, he may adopt the meme from another topically related external source; we estimate and subtract the impact of such external phenomena on adoption probability via the term $\int P_\text{a}(0,S) P(S|\kappa) dS$.

\subsection{Persistence of spreading}
We define the persistence of information diffusion as the ability of multiple exposures to boost information adoption. 
We introduce a new metric of spreading \textit{persistence}~\cite{Romero2011Differences} to capture this complex contagion effect. We define this persistence as the ratio between the average internal adoption probability at several exposures and the adoption probability at one exposure, namely $\text{Persistence} = \langle P^{\text{i}}_\text{a}(\kappa) \rangle_{\kappa>1}/P^{\text{i}}_\text{a}(\kappa=1)$, where $\langle \cdot \rangle$ is an average. Intuitively, our persistence metric captures if multiple exposures boost information diffusion in comparison with a single exposure.

\section{Impact of topicality on adoption probability}

Here, we characterize topical and non-topical mechanisms of information diffusion by studying the adoption probability versus social exposure and topical alignment. We analyze separately the probability of adoption caused by internal and external exposures. Among others, we show that the laws of topical information diffusion apply both to internal and external information diffusion.

\subsection{Topicality of memes vs adoption probability}
\label{sec:adoption-curves-memes}

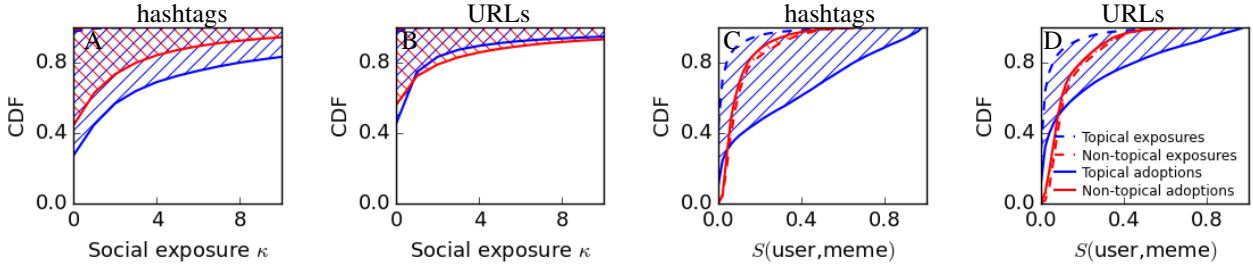
\begin{figure*}[tbh]
\begin{picture}(100,100)
\centering
\put(0,0){
\includegraphics[width=0.22\textwidth]{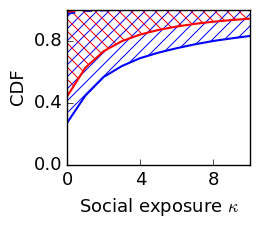}
\hspace{2mm}
\includegraphics[width=0.22\textwidth]{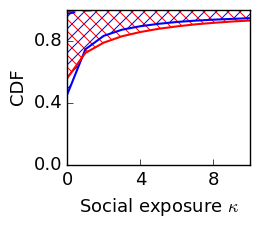}
\hspace{2mm}
\includegraphics[width=0.22\textwidth]{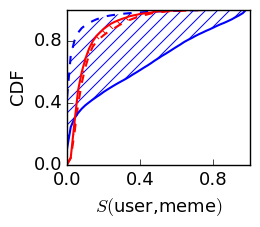}
\hspace{2mm}
\includegraphics[width=0.22\textwidth]{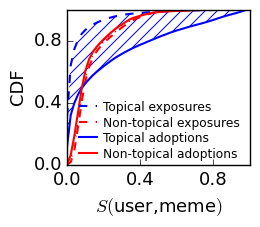}
}
\put(55,95){hashtags} \put(180,95){URLs} \put(300,95){hashtags} \put(420,95){URLs}
\put(35,85){A} \put(155,85){B} \put(277,85){C} \put(398,85){D}
\end{picture}
\caption{The cumulative distribution of: (A-B) social exposure $\kappa$ and (C-D) topical alignment $S(\text{user},\text{meme})$ of adoption events (solid line) and exposure events (dashed line) of hashtags and URLs. The blue lines correspond to topical memes, whereas the red lines correspond to non-topical memes. The filled area marks the difference between the distributions for adoption and exposure events.
}
\label{fig:distro_se}
\label{fig:distro_sim}
\end{figure*}

\begin{figure*}[tbh]
\begin{picture}(100,110)
\centering
\put(0,0){
\hspace{1mm}
\includegraphics[width=0.22\textwidth]{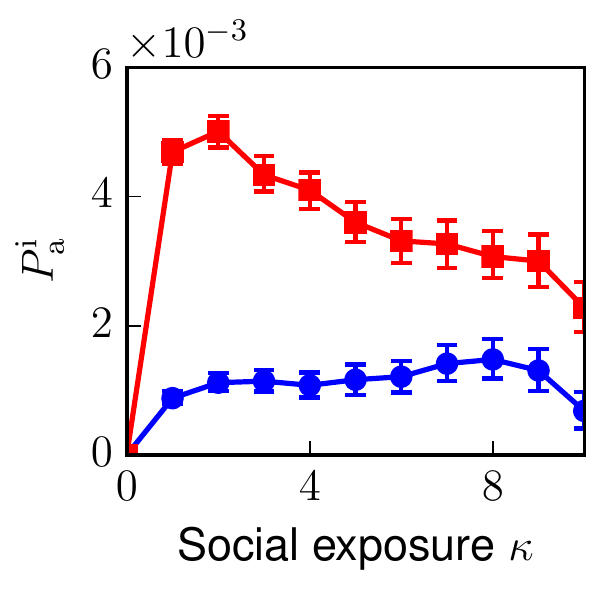}
\hspace{2mm}
\includegraphics[width=0.22\textwidth]{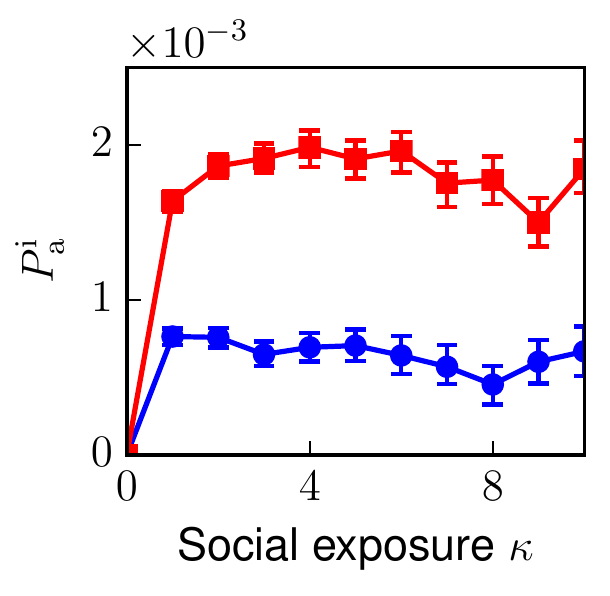}
\hspace{2mm}
\includegraphics[width=0.22\textwidth]{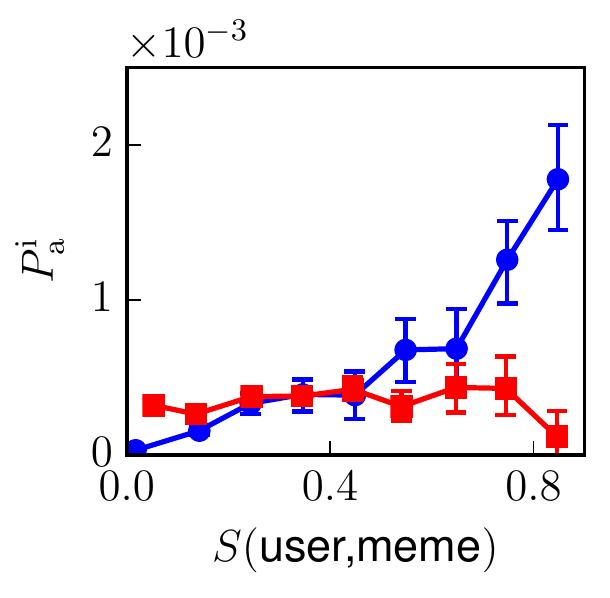}
\hspace{2mm}
\includegraphics[width=0.22\textwidth]{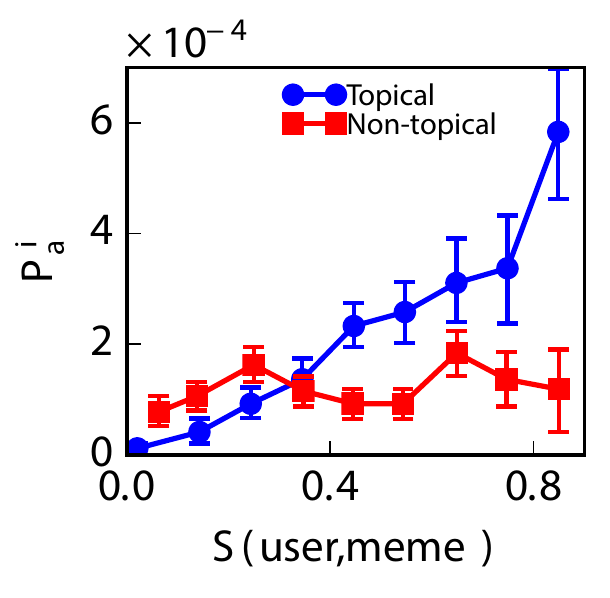}
}
\put(65,100){hashtags} \put(190,100){URLs} \put(310,100){hashtags} \put(430,100){URLs}
\put(33,88){A} \put(154,88){B} \put(277,88){C} \put(399,88){D}
\end{picture}
\caption{The adoption probability of hashtags and URLs as a function of: (A-B) social exposure $\kappa$, (C-D) topical alignment $S(\text{user},\text{meme})$ for topical (blue circles) and non-topical (red squares) memes. The error bars mark $95\%$ confidence intervals (from BCA bootstrap).}
\label{fig:adoption_curves}
\end{figure*}
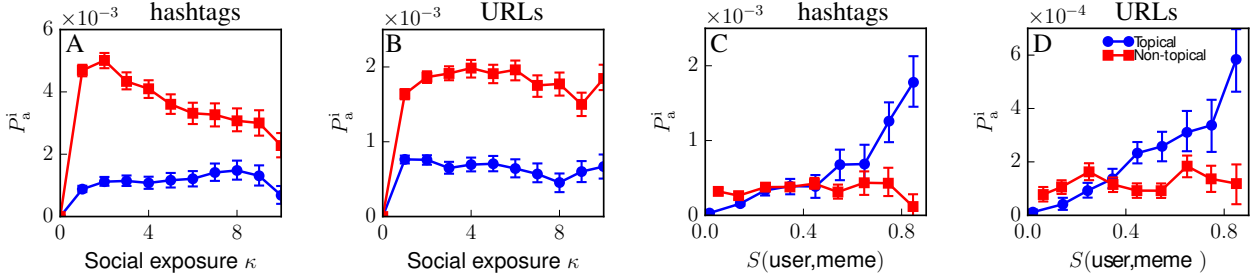

\begin{figure*}[tbh]
\begin{picture}(100,100)
\centering
\put(0,0){
\includegraphics[width=0.22\textwidth]{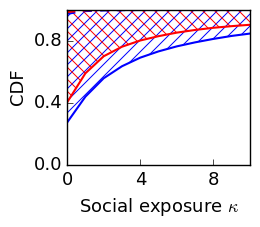}
\hspace{2mm}
\includegraphics[width=0.22\textwidth]{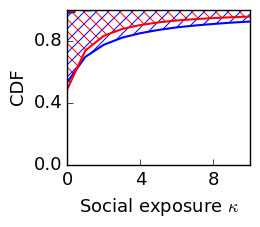}
\hspace{2mm}
\includegraphics[width=0.22\textwidth]{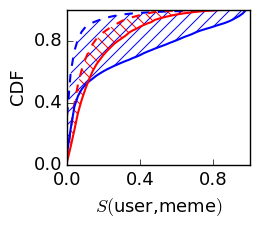}
\hspace{2mm}
\includegraphics[width=0.22\textwidth]{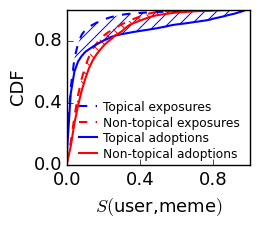}
}
\put(55,95){hashtags} \put(180,95){URLs} \put(300,95){hashtags} \put(420,95){URLs}
\put(35,85){A} \put(155,85){B} \put(277,85){C} \put(398,85){D}
\end{picture}
\caption{The cumulative distribution of: (A-B) social exposure $\kappa$ and (C-D) topical alignment $S(\text{user},\text{meme})$ of adoption events (line) and exposure events (dashed line) of hashtags and URLs. The blue lines correspond to topical users, whereas the red lines correspond to non-topical users.
}
\label{fig:distro_se_usertop}
\label{fig:distro_sim_usertop}
\end{figure*}
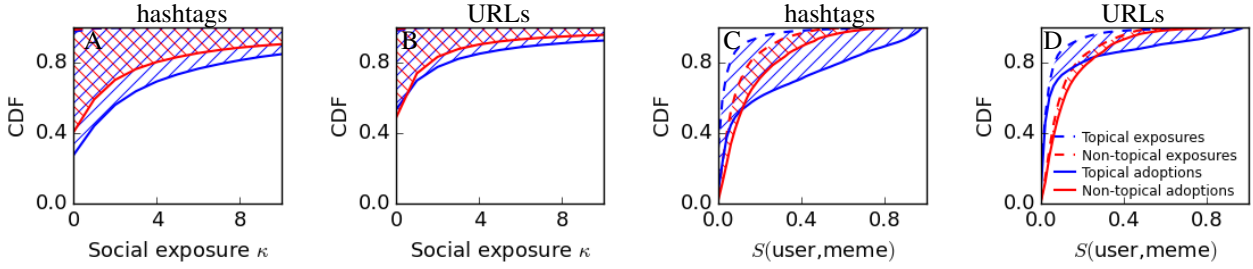

\begin{figure*}[tbh]
\begin{picture}(100,110)
\centering
\put(0,0){
\hspace{1mm}
\includegraphics[width=0.22\textwidth]{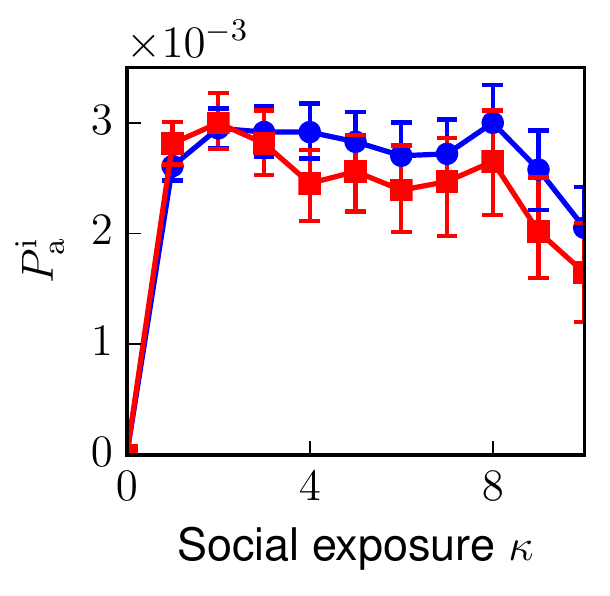}
\hspace{2mm}
\includegraphics[width=0.22\textwidth]{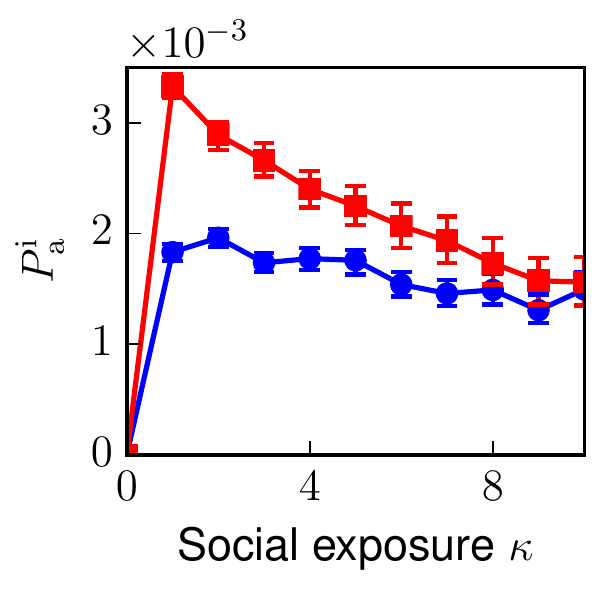}
\hspace{2mm}
\includegraphics[width=0.22\textwidth]{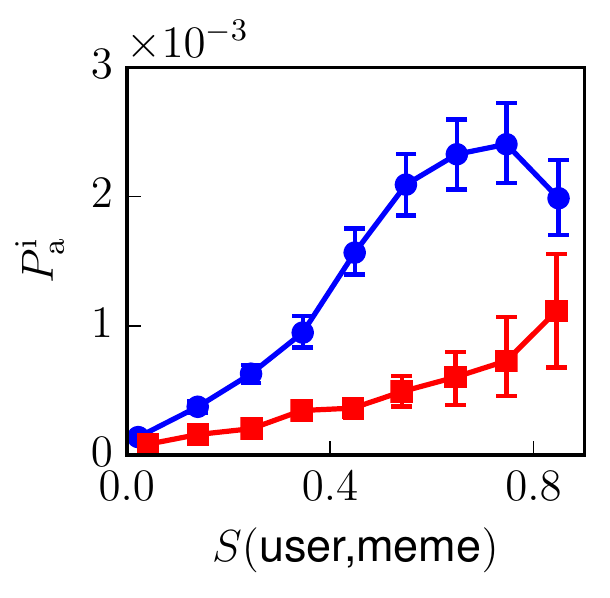}
\hspace{2mm}
\includegraphics[width=0.22\textwidth]{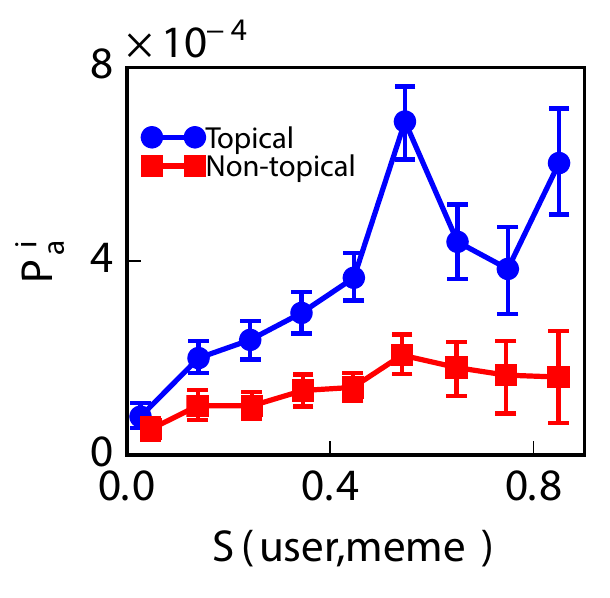}
}
\put(65,100){hashtags} \put(190,100){URLs} \put(310,100){hashtags} \put(430,100){URLs}
\put(33,88){A} \put(154,88){B} \put(277,88){C} \put(399,88){D}
\end{picture}
\caption{The adoption probability of hashtags and URLs as a function of: (A-B) social exposure $\kappa$, (C-D) topical alignment $S(\text{user},\text{meme})$ for topical (blue circles) and non-topical (red squares) users.}
\label{fig:adoption_curves_usertop}
\end{figure*}

We begin by measuring the distribution of social exposure $\kappa$ and topical alignment $S(\text{user},\text{meme})$ for adoption and exposure events of topical and non-topical memes (Figure~\ref{fig:distro_se}).
Prior works show that information cascades tend to be shallow and have multiple seed adopters who learn about the memes not from their friends, but from some external sources \cite{Goel2012structure}. Here, we confirm that over $30\%$ of hashtag adoptions and over $40\%$ of URL adoptions happened without a direct exposure through follower links (see y-value at $\kappa=0$ in Figures~\ref{fig:distro_se}A-B). 
Furthermore, we notice that the distribution of $S(\text{user},\text{meme})$ differs for topical and non-topical memes. In particular, the distributions of topical alignment $S(\text{user},\text{meme})$ are skewed to higher values for topical memes than non-topical memes. For instance, about $40\%$ of topical hashtags get adopted when the topical alignment is higher than $0.4$, whereas only $2\%$ of non-topical hashtags are adopted when the alignment is higher than $0.4$ (compare blue and red solid lines in Figure~\ref{fig:distro_se}C). Similar difference is present between topical and non-topical URLs (Figure~\ref{fig:distro_se}D). This signifies that topical memes spread differently than non-topical memes.

Also, there exists a large difference between the cumulative distributions of adoption and exposure events for topical memes (compare solid and dashed blue lines in Figure~\ref{fig:distro_se}). 
This difference can be quantified with Kolmogorov-Smirnov distance between the cumulative distributions, which is equivalent to the maximal vertical width of the corresponding filled area in Figure~\ref{fig:distro_se}. 
Because the K-S distance is high for topical memes both for social exposure $\kappa$ and topical alignment $S(\text{user},\text{meme})$, one can expect that both these properties can act as important discriminators for predicting whether an exposure event will lead to adoption or not.
By contrast, for non-topical memes the K-S distance between the distributions of the topical alignment is almost zero (red lines in Figures~\ref{fig:distro_se}C-D). This signifies that the topical alignment cannot predict the adoption of non-topical memes.

Next, we measure the probability of adoption via internal sources $P_\text{a}^{\text{i}}(\kappa)$, finding that it sharply increases with social exposure, reaching a maximum for hashtags and a maximum or a plateau for URLs (Figures~\ref{fig:adoption_curves}A-B). The probability decreases for higher number of exposures, because a user who does not adopt a meme after being exposed to it once or twice, will likely not adopt it, even if she is exposed to it by other people. Interestingly, the adoption probability is at least twice higher for non-topical memes than for topical memes. This indicates that non-topical memes are more viral and reach larger audiences than topical memes \cite{Weng2014Topicality}. In fact, non-topical hashtags have on average $934$ adopters and topical hashtags have $500$ adopters (significant difference, Mann-Whitney U test, $p=0.04$).
In other words, non-topical memes spread broadly, while topical memes diffuse more narrowly.

Finally, we measure the adoption probability as a function of the topical alignment between user's interests and the meme, namely $P_\text{a}^{\text{i}}(S(\text{user},\text{meme}))$. We find that the adoption probability curve has different shapes depending on meme's topicality (Figures~\ref{fig:adoption_curves}C-D). 
For topical memes, the adoption probability increases remarkably steadily with the topical alignment, both for hashtags and URLs. In stark contrast, the adoption probability remains flat for non-topical memes. 
To summarize, topical memes tend to diffuse narrowly among users who are interested in them, while non-topical memes spread broadly among all users.

\subsection{Topicality of users vs adoption probability}
\label{sec:adoption-curves-users}

Here, we describe the diffusion among users with topically narrow and topically diverse interests, i.e., among users with topical and non-topical interests~(Figures~\ref{fig:distro_se_usertop} and~\ref{fig:adoption_curves_usertop}). Overall, we observe the same tendencies as for topical and non-topical memes. However, these tendencies are less pronounced for topical and non-topical users. 
For instance, the adoption probability $P_\text{a}^{\text{i}}(S(\text{user},\text{meme}))$ is significantly higher for topical users than for non-topical users (Figures~\ref{fig:adoption_curves_usertop}C-D); however, here the adoption probability for non-topical users is not flat, but instead increases gradually.
Thus, in the remainder of this study, we focus on the results for topical and non-topical memes, because they show a cleaner distinction between the characteristics of topical and non-topical mechanisms of information diffusion.


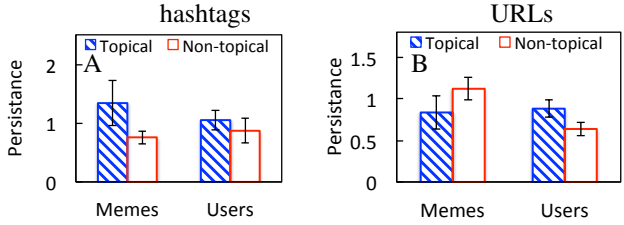
\begin{figure}[htp]
\begin{picture}(100,90)
\centering
\put(-5,0){
\includegraphics[width=0.225\textwidth]{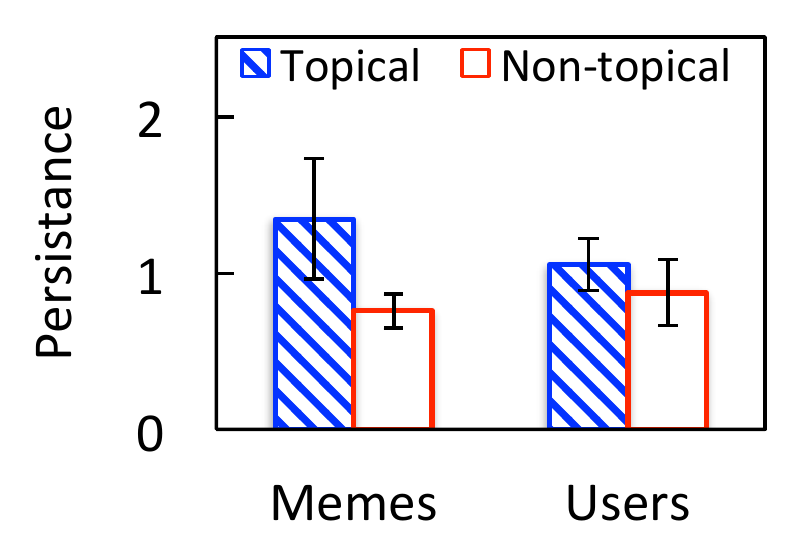}
\hspace{0mm}
\includegraphics[width=0.245\textwidth]{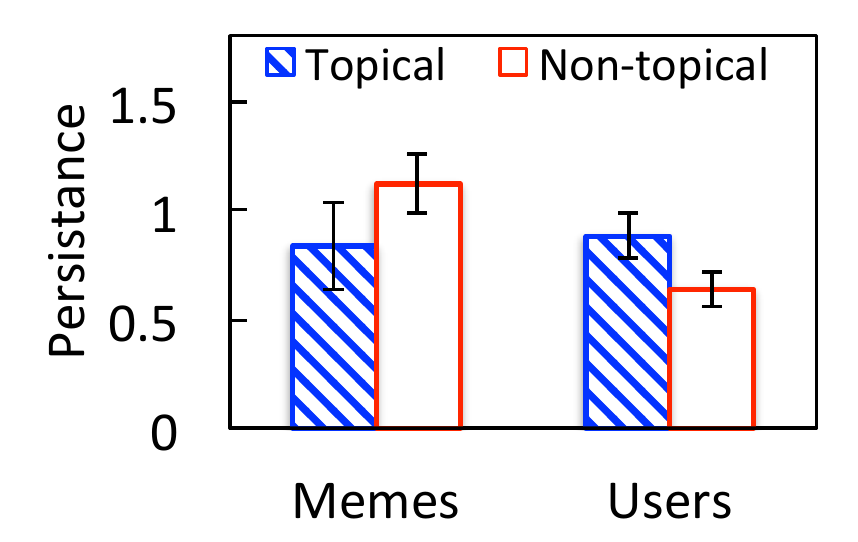}
}
\put(60,77){hashtags}\put(185,77){URLs}
\put(31,58){A}\put(155,58){B}
\end{picture}
\caption{Spreading persistence measured separately for topical and non-topical memes, as well as for topical and non-topical users. The errors bars are $95\%$ confidence intervals.}
\label{fig:persistence}
\end{figure}

Finally, we note that multiple social exposures to a meme boost topical diffusion more than non-topical diffusion. 
To this end, the adoption probability $P_\text{a}^{\text{i}}(\kappa)$ drops with social exposures more slowly for topical users than for non-topical users (Figures~\ref{fig:adoption_curves_usertop}A-B). 
This is much more evident for the spread of topical hashtags, for which the adoption probability increases with social exposure and reaches a maximum at eight exposures, much more than two social exposures needed to reach the maximum for non-topical hashtags (Figure~\ref{fig:adoption_curves}A). 
To our knowledge, it has not yet been observed that the maximum of adoption probability (aggregating over various hashtags) can be reached at such a high number of social exposures. 
Using our definition of spreading persistence, we find that the diffusion of topical hashtags is significantly boosted by multiple exposures (compare left bars of Figure~\ref{fig:persistence}A). One could expect this result, given the fact that many of topical hashtags contain novel, unproven, or controversial meaning. 
Furthermore, we find that the spread of URLs among users with topical interests is also significantly enhanced by multiple exposures (compare right bars of Figure~\ref{fig:persistence}B). We note here that the other two cases are insignificant due to the noise in the measurements of the adoption probability (evident in the Figure~\ref{fig:adoption_curves_usertop}D).
It is interesting to see that the complex contagion effect is present not only for topical memes, but also for users with topical interests. A plausible interpretation is that experts or enthusiasts of a given topical domain who decide to spread a certain information risk their topical reputation by doing so. Thus, they are more likely to diffuse information if multiple other people in their social network did so already. 

\vspace{1cm}

\subsection{Seed adopters and the probability of external adoption}
\label{sec:seeds}

Earlier in this section, we noted that a large fraction of users adopt memes without being exposed to these memes by any of their followees. We call such users \textit{seed adopters}. Seed adopters are the initiators of cascades propagating memes in a social network. Importantly, seed adopters at the moment of adoption are not influenced by social exposures internal to Twitter, but instead are influenced by external exposures.

We measure the adoption probability for seed adopters (described be Equation~\ref{eq:c}) by accounting only for adoption and exposure events at $\kappa=0$. As before, we split memes into topical and non-topical.
Essentially, we obtain the same shapes of the adoption probability curve for seed adopters as for all adopters (compare Figures~\ref{fig:adoption_curves_conditional}A-B against Figures~\ref{fig:adoption_curves}C-D). 
For topical memes the adoption probability increases with the topical alignment, whereas for non-topical memes it stays nearly constant. 
We present two explanations of this finding. First, a seed adopter may have stumbled on the meme through her sources external to Twitter, which likely are also homophilous and topically similar to that adopter. Second, the user may have actively searched for the memes that are topically aligned with her interests via search or by browsing profiles of other users or co-appearing hashtags. Note that this type of exposure can be also considered as external, because it happens via methods external to the follower network of Twitter; essentially, this is how we define external exposures in this study.
Finally, the comparison of figures reveals that the diffusion of URLs is nearly twice more intense via external sources rather than internal sources (the probability is twice higher in Figure~\ref{fig:adoption_curves_conditional}B than in Figure~\ref{fig:adoption_curves}D). By contrast, the diffusion of hashtags has similar intensity for external and internal sources. It is easy to interpret this result, knowing that URLs have the global scope of the Internet, whereas hashtags are specific to Twitter.\footnote{Twitter is the first platform where hashtags emerged, initially as a social convention in 2007, and then were implemented by Twitter as a part of the system in 2009.} Because of this distinction, the diffusion of URLs is about twice more intense via external exposures than internal exposures. Note that the diffusion of hashtags via ``external'' sources may correspond to the exposures that happen within Twitter via ways different that the follower network.

Furthermore, we find that in the domain of topical information diffusion seed adopters are relatively more topically aligned with the adopted memes than other adopters.
To compute the relative alignment we divide the alignment averaged over adoption events by the alignment averaged over exposure events, to account for homophily, namely $\langle S(\text{user},\text{meme}) \rangle_{\text{a}} / \langle S(\text{user},\text{meme}) \rangle_{\text{e}}$. We find that the relative topical alignment is significantly higher for seed adopters than for other adopters, both for topical hashtags and topical URLs (Figure~\ref{fig:seed}). Intuitively, seed adopters are more interested in the topic of a meme than other adopters, similar to early adopters being more enthusiastic about an innovation than other people \cite{Rogers1962Diffusion}. This sheds new light on the reason why memes have multiple seed adopters \cite{Goel2012structure}. Naturally, this effect is not present for non-topical memes.

\begin{figure}[t]
\begin{picture}(100,110)
\centering
\put(0,0){
\includegraphics[width=0.215\textwidth]{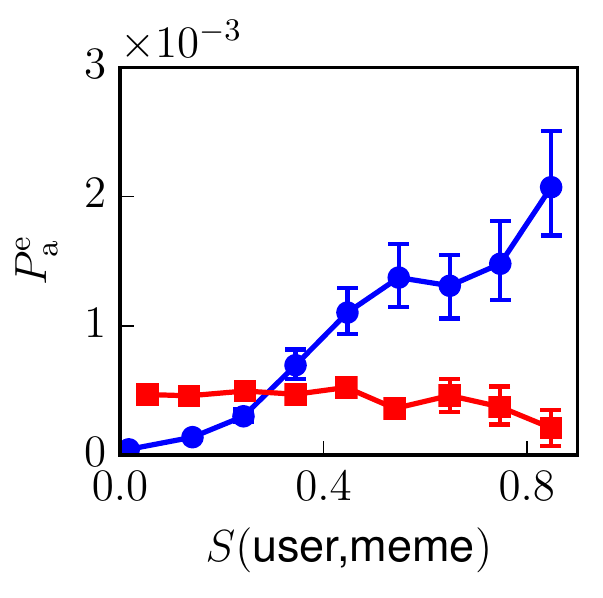}
\hspace{2mm}
\includegraphics[width=0.23\textwidth]{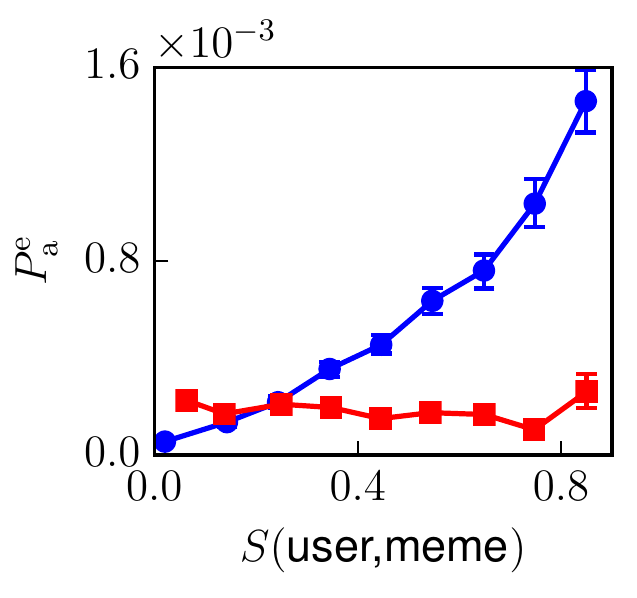}
}
\put(66,98){hashtags}\put(200,98){URLs}
\put(28,87){A}\put(156,87){B}
\end{picture}
\caption{The probability of external adoption ($\kappa=0$) as a function of topical alignment between a user and a meme for topical (blue) and non-topical (red) memes. The error bars mark $95\%$ confidence intervals.
}
\label{fig:adoption_curves_conditional}
\end{figure}
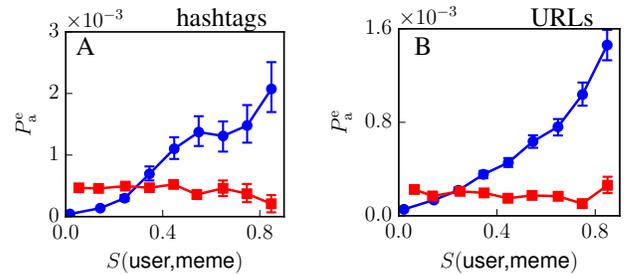

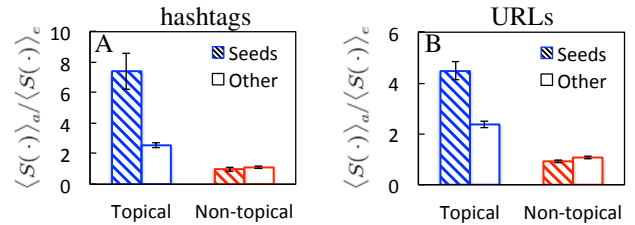
\begin{figure}[tp]
\begin{picture}(100,90)
\centering
\put(0,0){
\includegraphics[width=0.225\textwidth]{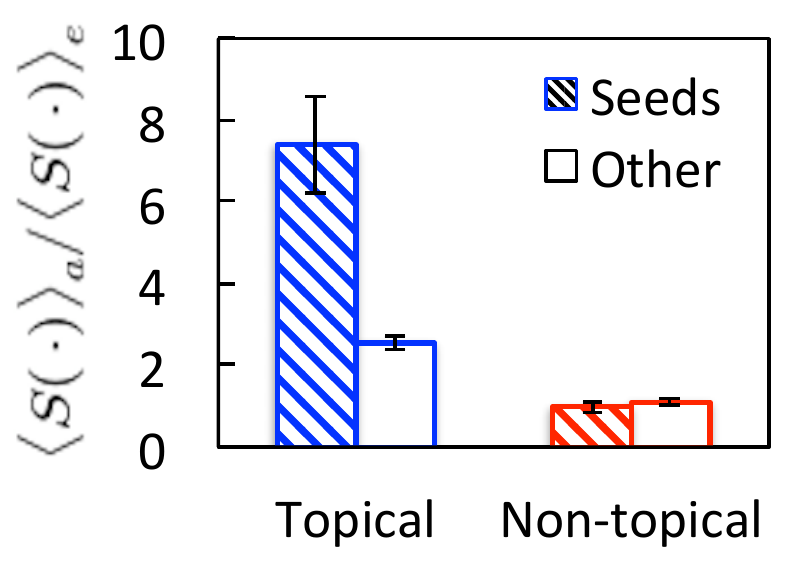}
\hspace{2mm}
\includegraphics[width=0.225\textwidth]{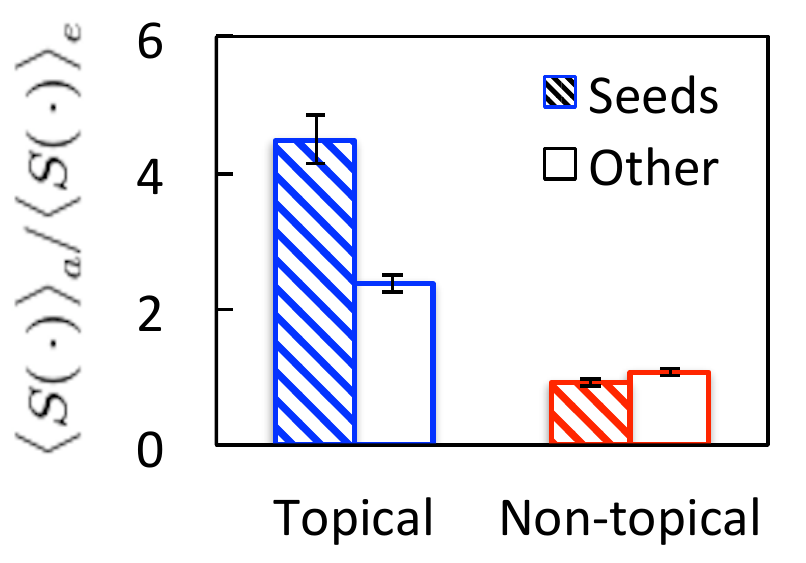}
}
\put(60,77){hashtags}\put(185,77){URLs}
\put(35,67){A}\put(160,67){B}
\end{picture}
\caption{The relative topical alignment for seed and other adopters. The error bars are $99\%$ confidence intervals. 
}
\label{fig:seed}
\end{figure}

\section{Discussion}
\label{sec:conclusions}

The findings presented in this study create the following schematic of topical information diffusion in Twitter. 
Topical memes start to spread from the enthusiasts or experts of the corresponding topic. 
Then, they diffuse narrowly to other users whose interests are topically aligned with the topics of the memes.
Multiple exposures to the same meme tend to boost topical information diffusion, as predicted by the theory of complex contagion, since many topical memes contain novel or controversial information.
Restarts of information cascades are frequent and are initiated by seed adopters who tend to be relatively more topically aligned with the given information than other adopters. 
Overall, we have shown that topical information diffusion differs from non-topical information diffusion in several meaningful ways.

Our work explores the gap between the studies of information diffusion and recommender systems by showing that users even without personalized recommendations adopt information that is topically related to them.
Clearly, recommender systems may strengthen or weaken the natural human tendency to spread topically aligned information.
How recommender system affect information diffusion? Could we define a recommender system that facilitates information diffusion without causing a filtering bubble? This is a set of questions that could be addressed in future studies.
Furthermore, in this study, rather than proposing and training a supervised model predicting information diffusion, we use a state-of-the-art unsupervised topic modeling method and analyze the impact of topicality on information diffusion independently.
As such, our results provide an empirical lower estimate of the impact of topicality on diffusion. Future studies could propose a topic model merged with a diffusion model as an effective recommender system. 

To conclude, we note that information diffusion is a fundamental building block of many collective processes happening in social networks, e.g., viral marketing, spreading of political movements and innovation, and opinion formation. Thus, our findings have wide implications for understanding, predicting, and controlling such processes.

\subsection*{Acknowledgments}
We thank Komal Agrawal for her initial work on this project, Robert West for his valuable suggestions, Vlad Niculae for his insights into the classification problem, and Vlad Niculae, Manuel Gomez-Rodriguez, and Juhi Kulshrestha for their feedback on early versions of this study. Finally, we thank the anonymous reviewers for helping to improve this work.
Also, we acknowledge the support of project IMPECS ``Understanding, Leveraging and Deploying Online Social Networks''.

\bibliographystyle{aaai}
\bibliography{jabref.bib}

\end{document}